\documentclass[lettersize,journal]{IEEEtran}
\usepackage{amsmath,amsfonts}
\usepackage{algorithmic}
\usepackage{algorithm}
\usepackage{array}
\usepackage[caption=false,font=normalsize,labelfont=sf,textfont=sf]{subfig}
\usepackage{textcomp}
\usepackage{stfloats}
\usepackage{url}
\usepackage{verbatim}
\usepackage{graphicx}
\usepackage{cite}

\usepackage{multirow}
\begin{document}

\title{Deep Joint CSI Feedback and Multiuser Precoding for MIMO OFDM Systems}

\author{Yiran Guo, Wei Chen,~\IEEEmembership{Senior Member,~IEEE,} Jialong Xu,~\IEEEmembership{Member,~IEEE,} Lun Li, Bo Ai,~\IEEEmembership{Fellow,~IEEE}

\thanks{Yiran Guo, Wei Chen, Jialong Xu and Bo Ai are with School of Electronic and Information Engineering, Beijing Jiaotong University, Beijing, China. Lun Li is with the State Key Laboratory of Mobile Network and Mobile Multimedia, Shenzhen, China. This work was supported by ZTE Industry-University-Institute Cooperation Funds under Grant No. HC-CN-20221208006.}
}


\maketitle

\begin{abstract}
The design of precoding plays a crucial role in achieving a high downlink sum-rate in multiuser multiple-input multiple-output (MIMO) orthogonal frequency-division multiplexing (OFDM) systems. In this correspondence, we propose a deep learning based joint CSI feedback and multiuser precoding method in frequency division duplex systems, aiming at maximizing the downlink sum-rate performance in an end-to-end manner. Specifically, the eigenvectors of the CSI matrix are compressed using deep joint source-channel coding techniques. This compression method enhances the resilience of the feedback CSI information against degradation in the feedback channel. A joint multiuser precoding module and a power allocation module are designed to adjust the precoding direction and the precoding power for users based on the feedback CSI information. Experimental results demonstrate that the downlink sum-rate can be significantly improved by using the proposed method, especially in scenarios with low signal-to-noise ratio and low feedback overhead. 


\end{abstract}


\section{Introduction}

Massive multiple-input multiple-output (MIMO) significantly improves spectrum efficiency and system throughput by exploiting spatial domain resources. Its successor, i.e., ultra-massive MIMO, is expected to become one of the key technologies in 6G to further boost the transmit rate and support emerging applications, e.g., metaverse and extended reality (XR). The performance of MIMO relies on accurate channel state information (CSI). In frequency division duplex (FDD) systems, the downlink CSI needs to be fed back from the user equipments (UEs) to the base station (BS). In this situation, CSI compression is preferred to reduce the overhead of CSI carried on the uplink resources and improve the uplink transmission efficiency. After the BS reconstructs the downlink CSI through the compressed information, precoding is performed to utilize MIMO channel characteristics and improve the system throughput.

The overhead and computational complexity of traditional CSI feedback methods, such as codebook-based methods and compression sensing-based methods, will be increasingly challenged as the number of antennas and users grows \cite{CSI_overview}. Recently, deep learning (DL) has shown great potential in CSI compression \cite{CsiNet,CsiNet_plus,CsiFormer}, where CSI feedback is treated as a source compression task. Moreover, the eigenvector extracted from CSI can be compressed via DL to further alleviate feedback overhead, which has been employed as the precoding vector for single-user MIMO in \cite{Implicit_CSI_Feedback}. These CSI compression and feedback methods are based on the paradigm of separate source-channel coding (SSCC) and assume perfect feedback channel. Thus, they are susceptible to the ``cliff effect''. That is, if the actual feedback channel conditions fall below expectations, the reconstruction quality of the CSI declines abruptly. Following the paradigm of joint source-channel coding, a deep joint source-channel coding (DJSCC) network for CSI feedback is proposed to relieve the ``cliff effect'' and improve the CSI reconstruction quality \cite{DJSCC}. 

To support multiple users, the CSI information is exploited in multiuser (MU) precoding. DL based algorithms show lower computational complexity compared to traditional iteration based algorithms, e.g., weighted minimum mean square error (WMMSE) algorithm, for MU precoding \cite{MIMO_Pre}. Since the precoding performance relies on the reconstruction accuracy of CSI feedback, a DL based joint CSI feedback and precoding method is proposed for relieving the impact of CSI feedback error \cite{MU_CSI_Pre}. Furthermore, considering that CSI feedback error is also relevant to the channel estimation accuracy, an end-to-end network integrating the channel estimation, the CSI feedback and the precoding design is proposed to reveal remarkable precoding performance \cite{MU_WeiYu}. However, none of the aforementioned methods consider the use of eigenvectors with low feedback overhead for precoding design in MIMO-orthogonal frequency-division multiplexing (OFDM) systems as the bandwidth and the number of antennas increase. 

In this correspondence, we propose a DJSCC-based joint task-oriented CSI feedback and precoding network (JFPNet) for MU-MIMO-OFDM systems. Following \cite{CsiNet,CsiNet_plus,CsiFormer,MRFNet,Implicit_CSI_Feedback,DJSCC}, we consider perfect CSI at the UE side. By applying singular value decomposition (SVD), Each UE obtains the eigenvector matrix, which is fed into the designed DJSCC encoder. After the wireless transmission, the received noisy feedback symbols are decoded by the DJSCC based decoder at the BS. Then a joint MU precoding module and a power allocation module are designed to predict the precoding matrix at the BS. It is worth noting that the DJSCC decoder intends to further extract useful CSI information relevant to the precoding task from the noisy feedback, rather than recovering the source information, i.e., the eigenvector (matrix). To maximize the downlink sum-rate, the proposed network is optimized in an end-to-end manner.

The rest of this correspondence is organized as follows. Section \ref{section1} introduces the system model in FDD MU-MIMO-OFDM systems. The proposed framework including the joint multiuser precoding module, the DL based power allocation module, and the JFPNet is introduced in Section \ref{section2}. In Section \ref{section3} presents the experimental results and Section \ref{section4} concludes our work.

\textit{Notations:}In this correspondence, we denote Vectors and Matrices with boldface lower- and upper-case letters, respectively. ${\mathbf{A}}^H$ and ${\mathbf{A}}^T$ are complex conjugate transpose of ${\mathbf{A}}$ and transpose operation of ${\mathbf{A}}$, respectively. ${\left\| \cdot \right\|_2}$, ${\left\| \cdot \right\|_F}$ and $\left| \cdot \right|$ represent the $\ell\text{-}2$ norm, the Frobenius norm and matrix determinant, respectively. $\mathbb{E}\left(\cdot \right)$ denotes the statistical expectation. $\mathbf{I}_c$  is a $c$-dimensional identity matrix.

\section{System model}
\label{section1}

We consider an FDD MU-MIMO-OFDM system with ${N_d}$ subcarriers. The BS is equipped with ${N_t}$ antennas and serves $K$ users with ${N_r}$ antennas for each. ${\mathbf{H}_{k}} \in {\mathbb{C}^{{N_d} \times {N_r} \times {N_t} }}$ denotes the downlink channel matrix between the BS and the user ${k}$, also called the full-CSI. Following the 5G NR standards for subband implicit feedback, we initially partition the full CSI into Resource Blocks (RBs) for assessing the downlink sum-rate within each RB. Subsequently, we aggregate multiple consecutive RBs to form a subband and develop shared precoding vectors for each subband \cite{Implicit_CSI_Feedback,3GPP38214}. Every $a$ consecutive subcarriers form a resource block (RB) and every $b$ RBs forms a subband, which leads to $N_{RB}$ RBs and ${N_b}$ subbands, i.e., $N_{d}=a N_{RB}$ and $N_{RB}=b N_b$.

In the FDD system, it is necessary to feed back the downlink CSI explicitly or implicitly from UEs to the BS to support the precoding design. To avoid excessive CSI feedback overhead,  the spatial-frequency domain CSI at each UE is initially processed, which can be expressed as 
\begin{equation}
	\label{F_preprocess}
	\left\{{\mathbf{M}_k,{\mathbf{E}}_k,{\mathbf{H}_{dl,k}}}\right\} = \mathcal{P}({\mathbf{H}_k}).
\end{equation}
${\mathbf{M}_k} = \left[ {{{\mathbf{M}}_{k}^{\left({1}\right)}}, \cdots ,{{\mathbf{M}}_{k}^{\left({N_b}\right)}}} \right] \in {{\mathbb{C}}^{{N_b} \times {N_r} \times {N_s}}}$ 
represents the eigenvector matrix of the $k$-th user, and ${N_s}$ denotes the number of data streams, which satisfies ${N_s} \le {N_r}$. ${\mathbf{M}}_{k}^{\left({n_b}\right)}=\left[ {{{\mathbf{m}}_{k,1}^{\left({n_b}\right)}}, \cdots ,{{\mathbf{m}}_{k,{N_{s}}}^{\left({n_b}\right)}}} \right]$ (${n_b} = 1, \ldots ,{N_b}$) should satisfy the power constraint ${\left\| {{\mathbf{m}_{k,n_s}^{\left({n_b}\right)}}} \right\|_2^2} = 1$ (${n_s} = 1, \ldots ,{N_s}$).  ${{\mathbf{E}}_k} = \left[ {{{\mathbf{E}}_{k}^{\left({1}\right)}}, \cdots ,{{\mathbf{E}}_{k}^{\left({N_b}\right)}}} \right] \in {{\mathbb{R}}^{{N_b} \times {N_s} \times {N_s}}}$ is the eigenvalue matrix of the $k$-th user, and the diagonal matrix ${{{\mathbf{E}}}_{k}^{\left({n_b}\right)}}$ (${n_b} = 1, \ldots ,{N_b}$) is constructed by the square of the eigenvalue of each data stream for the ${n_b}$-th subband. 
In BS, a subcarrier is allocated within each RB for downlink CSI acquisition through the transmission of a reference signal. The RB-level downlink CSI obtained from $N_{RB}$ RBs is denoted as ${\mathbf{H}_{dl,k}}= \left[ {{{\mathbf{H}}_{dl,k}^{\left(1\right)}}, \cdots,{{\mathbf{H}}_{dl,k}^{\left(N_{RB}\right)}}} \right] \in \mathbb{C}^{N_{RB} \times N_r \times N_t}$, which is used for calculating the downlink sum-rate. $\mathcal{P}(\cdot)$ represents the preprocessing function.

The eigenvector matrix $\mathbf{M}_k$ is compressed and fed back from the $k$-user to the BS in the manner of DJSCC, which can be expressed as
\begin{equation}
	\label{DJSCC}
	{{\hat{\mathbf{M}}}_k} = {\mathcal{D}_{\beta }}\left( {{\mathcal{C}}\left( {{{\mathcal{E}}_{\alpha}}\left( {{\mathbf{M}_k}} \right),{{\mathbf{H}}_{ul,k}},\sigma_{ul}^{2}} \right)} \right) \in \mathbb{C}^{N_b \times N_r \times N_s}.
\end{equation}
${{{\mathcal{E}}_{{\alpha}}}(\cdot)}$ denotes the DJSCC encoder with the parameter set ${\alpha}$, and its output is $\mathbf{s}_k \in \mathbb{C}^n$. ${{{\mathcal{C}}}(\cdot)}$ consists of preprocessing the $\mathbf{s}_k$ at the UE, transmitting the preprocessed signals through the uplink channel, and acquiring the detected symbols $\hat{\mathbf{s}}_k \in \mathbb{C}^n$ from postprocessing the received signals at the BS. Then ${{\hat{\mathbf{M}}}_k}$ is decoded by the DJSCC decoder ${{{\mathcal{D}}_{{\beta}}}(\cdot)}$ with the parameter set ${\beta}$ from $\hat{\mathbf{s}}_k$. The uplink channel matrix and the power of additional white Gaussian noise (AWGN) are represented by ${\mathbf{H}_{ul,k}}\in \mathbb{C}^{N_u \times N_t \times N_r}$ and $\sigma_{ul}^{2}$, respectively. $N_u$ is the subcarrier number assigned for uplink transmission of $\mathbf{s}_{k}$.

Compared with huge resources reserved for the eigenvector matrix transmission, much less transmission resource is employed for transmitting the eigenvalue matrix $\mathbf{E}_k$. Here, we assume perfect feedback for eigenvalue matrix transmission. The decoded eigenvector matrices and the perfect eigenvalue matrices of $K$ users are aggregated at the BS as ${{\mathbf{{\hat M}}}}=\left\{{{{\mathbf{{\hat M}}}_1},\cdots,{{\mathbf{{\hat M}}}_K}}\right\}$ and ${{\mathbf{{E}}}}=\left\{{{{\mathbf{{E}}}_1},\cdots,{{\mathbf{{E}}}_K}}\right\}$, respectively. Then the precoding matrix for each user can be obtained via network ${{{\mathcal{G}}_{{ \theta }}}(\cdot)}$ with parameter sets $\theta$,
\begin{equation}
	\label{BS_net}
	\mathbf{O} = {\mathcal{G}_{{ \theta }}}\left( {{ {\mathbf{\hat M}}},{{\mathbf{E}}}} \right),
\end{equation}
where $\mathbf{O}=\left\{{{\mathbf{O}_1},\cdots,{\mathbf{O}_K}}\right\}$ is the set of $K$ uesrs' precoding metrics. $\mathbf{O}_k=\left[{{\mathbf{O}_k^{\left({1}\right)}},\cdots,{\mathbf{O}_k^{\left({N_b}\right)}}}\right]\in {\mathbb{C}^{N_b \times N_t \times N_s}}$ is the precoding matrices for the $k$-th user.

After the precoding process, the $k$-th user's received signals at the channel $\mathbf{H}_{dl,k}^{\left({n_{rb}}\right)}$ can be expressed as
\begin{equation}\label{receive_signal}
\begin{split}
{\bf{y}}_k^{\left( {{n_{rb}}} \right)} = {\bf{W}}_k^{\left( {{n_{rb}}} \right)}\Bigg(& \underbrace {{\bf{H}}_{dl,k}^{\left( {{n_{rb}}} \right)}{\bf{O}}_k^{\left( \gamma  \right)}{\bf{x}}_k^{\left( {{n_{rb}}} \right)}}_{{\rm{desired \; signal}}} +\\
& \underbrace {\sum\limits_{m \ne k} {{\bf{H}}_{dl,k}^{\left( {{n_{rb}}} \right)}{\bf{O}}_m^{\left( \gamma  \right)}{\bf{x}}_m^{\left( {{n_{rb}}} \right)}} }_{{\rm{interference \; from\;other \; users}}}+ \underbrace {{{\bf{n}}^{\left( {{n_{rb}}} \right)}}}_{{\rm{noise}}} \Bigg),
\end{split}
\end{equation}
where $\gamma  = \left\lfloor {\frac{{{n_{rb}} - 1}}{a} + 1} \right\rfloor$ is the subband index so that RBs in a subband share the same precoding scheme \cite{Implicit_CSI_Feedback}. ${{{\mathbf{n}}^{\left( {{n_{rb}}} \right)}}}$ is the AWGN noise with zero mean and variance $\sigma _{dl}^2$. ${{\mathbf{x}}_k^{\left( {{n_{rb}}} \right)}} \in \mathbb{C}^{N_s}$ is the message transmitted from the BS to the $k$-th user passing through the channel ${{\mathbf{H}}_{dl,k}^{\left(N_{RB}\right)}}$. The transmitted message satisfies the power constraint $\frac{1}{{{N_s}}}{\mathbb{E}}\Big( {\big\| {{\mathbf{x}}_k^{\left( {{n_{rb}}} \right)}} \big\|_2^2} \Big) = 1$. ${\mathbf{W}}_k^{\left( {{n_{rb}}} \right)} \in {\mathbb{C}^{{N_s} \times {N_r}}}$ denotes the combining matrix for the ${n_{rb}}$-th RB of the $k$-th user. The target of our design is to maximize the downlink average sum-rate in full band, which is expressed as
\begin{equation}
	\label{R_object}
	\mathop {{\rm{maximize}}}\limits_{\alpha,\beta,\theta} \frac{1}{{{N_{RB}}}}\sum\limits_{{n_{rb}} = 1}^{{N_{RB}}} {\sum\limits_{k = 1}^K {R_k^{\left( {{n_{rb}}} \right)}} },
\end{equation}
where $R_k^{\left( {{n_{rb}}} \right)}$ denotes the downlink achievable rate of the $k$-th user in the $n_{rb}$-th RB and its expression is given in Eq.~\eqref{R_k}.

\begin{figure*}[ht] 
\centering 
\begin{equation}
\label{R_k}
R_k^{\left( {{n_{rb}}} \right)} = {\log _2}\left| {{{\bf{I}}_{{N_s}}} + \frac{{{\bf{W}}_k^{\left( {{n_{rb}}} \right)}{\bf{H}}_{dl,k}^{\left( {{n_{rb}}} \right)}{\bf{O}}_k^{\left( \gamma  \right)}{{\left( {{\bf{W}}_k^{\left( {{n_{rb}}} \right)}{\bf{H}}_{dl,k}^{\left( {{n_{rb}}} \right)}{\bf{O}}_k^{\left( \gamma  \right)}} \right)}^H}}}{{\sum\limits_{m \ne k} {{\bf{W}}_k^{\left( {{n_{rb}}} \right)}{\bf{H}}_{dl,k}^{\left( {{n_{rb}}} \right)}{\bf{O}}_m^{\left( \gamma  \right)}{{\left( {{\bf{W}}_k^{\left( {{n_{rb}}} \right)}{\bf{H}}_{dl,k}^{\left( {{n_{rb}}} \right)}{\bf{O}}_m^{\left( \gamma  \right)}} \right)}^H} + \sigma _{dl}^2{\bf{W}}_k^{\left( {{n_{rb}}} \right)}{{\left( {{\bf{W}}_k^{\left( {{n_{rb}}} \right)}} \right)}^H}} }}} \right|
\end{equation}
\hrulefill 
\vspace*{8pt} 
\end{figure*}

\section{DJSCC-based joint task-oriented CSI feedback and precoding network}
\label{section2}
\begin{figure*}[!t]
	\centering
	\includegraphics[width=6.9in]{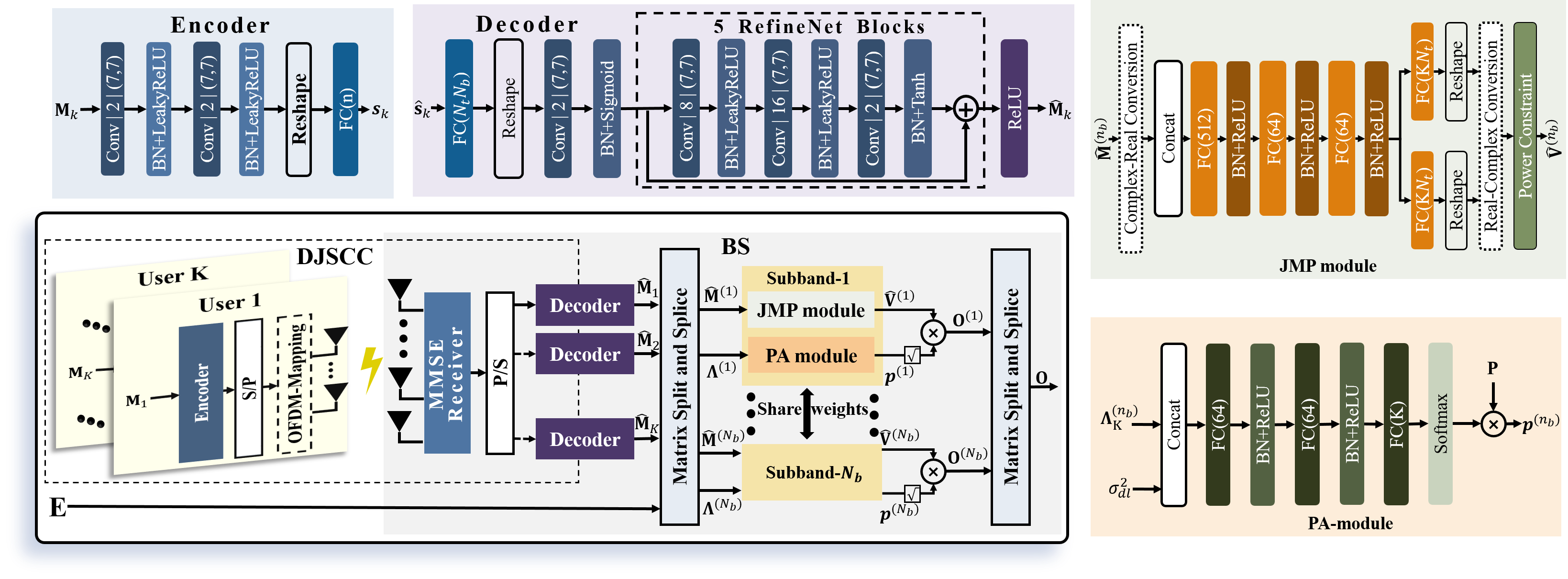}
	\caption{Architecture of the JFPNet. Data processing parts without training parameters are omitted from the figure. $\mathbf{O}^{\left({n_b}\right)}=\left\{{\mathbf{O}_1^{\left({n_b}\right)},\cdots,\mathbf{O}_k^{\left({n_b}\right)}}\right\}$ ($n_b=1,\ldots,N_b$) denotes the set of $K$ users' precoding vectors in $n_b$-th subband.}
	\label{end_to_end_MU_joint_CSI_FB_and_precoding}
\end{figure*}



In this section, we propose a DL-based network that implicitly feeds back the extracted CSI information, i.e., the eigenvector matrix, in a DJSCC manner, and generates the precoding matrix by adjusting the precoding direction and the precoding power for multiple users. 

To simplify the description, we will consider the case $N_s=1$ without loss of generality. As shown in Fig.~\ref{end_to_end_MU_joint_CSI_FB_and_precoding}, each UE compresses its eigenvector matrix by the DJSCC encoder. After performing the series-parallel transformation and OFDM mapping, which are mentioned as the preprocessing in section \ref{section1}, the compressed symbols are transmitted to the BS through the fading channel. At the BS, the received symbols of $N_t$ antennas are detected by the maximum ratio combining (MRC) and parallel-serial transformation, which represents the postprocessing in section \ref{section1}. Then the received symbols are decoded by the DJSCC decoder. After processing the feedback from $K$ UEs, the decoded multiuser CSI features are fed forward and processed for designing the precoding matrix for MU communication. Note that the DJSCC encoder of all UEs share the same set of parameters. Architectures of the DJSCC encoder and the DJSCC decoder are illustrated at the top-left of Fig.~\ref{end_to_end_MU_joint_CSI_FB_and_precoding}. The notation ``$C|(F, F)$'' in the convolutional layer denotes $C$ filters of the size $F$. 

We create precoding vectors for multiple users within the same subband, assuming no inter-subband interference for simplicity. At the BS, the eigenvector matrix set $\hat{\mathbf{M}}$ reconstructed by DJSCC decoder and eigenvalue matrix set $\mathbf{E}$ are first divided into different subband sets $\hat{\mathbf{M}}^{\left({n_b}\right)}$ ${\left({n_b=1,\ldots, N_b}\right)}$ and $\mathbf{\Lambda}^{\left({n_b}\right)}$ ${\left({n_b=1,\ldots,N_b}\right)}$ by matrix splitting and splicing, where $\hat{\mathbf{M}}^{\left({n_b}\right)}=\left\{{{\hat{\mathbf{M}}_1^{\left({n_b}\right)}},\cdots,\hat{\mathbf{M}}_K^{\left({n_b}\right)}}\right\}$ and $\mathbf{\Lambda}^{\left({n_b}\right)}=\left\{{{\mathbf{\Lambda}_1^{\left({n_b}\right)}},\cdots,\mathbf{\Lambda}_K^{\left({n_b}\right)}}\right\}$. 

For each subband, we design a joint multiuser precoding (JMP) module and a power allocation (PA) module to adjust the direction and the power of the precoding vector, respectively. The precoding vectors are generated by multiplying the output of the two modules. Then $N_B$ subband precoding matrices are reshaped by the matrix splitting and splicing module to form the precoding matrices for different users. The same network architecture and parameters is reused for all subbands. In the sequel, we introduce the subband precoding network, including the JMP module and the PA module.

\subsection{Joint Multiuser Precoding Module}
Based on the decoded CSI features, the JMP module is designed to indicate the precoding directions for UEs, expressed as
\begin{equation}
	\label{joint_precoding}
	\hat{\mathbf{V}}^{\left({n_b}\right)} = {\mathcal{J}}\left( {{\hat {\mathbf{M}}^{\left({n_b}\right)}}} \right),
\end{equation}
where ${\mathbf{{\hat V}}}^{\left({n_b}\right)}=\left\{{{\mathbf{{\hat v}}_1}^{\left({n_b}\right)},\cdots,{\mathbf{{\hat v}}_K}^{\left({n_b}\right)}}\right\} $ is the set of the $n_b$-th subband's precoding vector for $K$ users and satisfies the power constrain: ${\Big\| {\hat {\mathbf{v}}_k^{\left({n_b}\right)}} \Big\|_2^2} = 1$ $\left({k=1,\ldots,K}\right)$, where ${\hat {\mathbf{v}}_k^{\left({n_b}\right)}}\in \mathbb{C}^{N_t}$.

The network architecture of the JMP module on the $n_b$-th subband is shown in the top-right part of Fig.~\ref{end_to_end_MU_joint_CSI_FB_and_precoding}. Akin to \cite{MU_WeiYu}, the proposed JMP module adopts an FC based network to extract the precoding direction from multiuser's feedback CSI information. Specifically, the real and imaginary parts of the eigenvector for every user under $n_b$-th subband, reconstructed through feedback, are initially concatenated into a real vector. Subsequently, an FC network with normalization layers is employed to extract the correlation among users. Then the real and imaginary parts of the MU precoding vectors under the $n_b$-th subband are generated by the FC layer and the reshape layer. Finally, the MU precoding vectors under $N_b$ subbands are obtained by the real-complex conversion and the power constraint.

\subsection{Power Allocation Module}

At the BS, the eigenvalues of different users within the same subband, along with the noise power $\sigma_{dl}^2$ of the downlink channel, are simultaneously fed into the PA module to allocate appropriate transmit power for each UE, given as
\begin{equation}
	\label{PA_module}
	{\mathbf{p}^{\left({n_b}\right)}} = {\mathcal{F}}\left({\mathbf{\Lambda}^{\left({n_b}\right)},{\sigma_{dl} ^2},P} \right).
\end{equation}
where $P$ is the total power, ${\mathbf{p}^{\left({n_b}\right)}} \in {{\mathbb{R}}^{K}}$ is the power allocation vector of the ${n_b}$-th subband. The power allocation matrix satisfied the sum power constraint $\mathbb{E}\left( {\left\| {{\mathbf{p}^{\left({n_b}\right)}}} \right\|_2^2} \right) \le P$. 
 


The network architecture of the PA module within the ${n_b}$-th subband is shown in the bottom-right of Fig.~\ref{end_to_end_MU_joint_CSI_FB_and_precoding}. Similar to the JMP module, the PA module is constructed by alternatively stacking FC layers, batch normalization layers and ReLU layers. Normalization is achieved by applying the softmax activation before generating the output. In the end, the results are scaled by the total power $P$ in order to distribute power while still adhering to the total power limitations.


\section{Experimental results}
\label{section3}
In this section, we will assess the proposed method in comparison to existing methods, evaluate the effectiveness of the proposed modules through an ablation study, and examine how the feedback overhead impacts the performance of the proposed method.

The urban macrocell (Uma) with an uplink center frequency 2.1 GHz and a downlink center frequency 1.9 GHz is considered in the following experiments. Both the uplink bandwidth and the downlink bandwidth are set to 10 MHz. The uplink CSI and the downlink CSI are generated by QuaDRiGa \cite{QuaDRiGa}, following the 3rd Generation Partnership Project (3GPP) TR 38.901 \cite{3GPP38901}. The number of subcarriers $N_{f_{dl}}=624$ is adopted, corresponding to 52 RB and 13 subbands, which conforms to the relationship among subcarrier, RB, and subband defined in 5G standard \cite{3GPP38214}. Training, validation, and test datasets contain 100,000, 30,000, and 20,000 sample pairs, respectively. During the training stage, Adam optimizer with an initial learning rate of 0.001 is chosen to optimize the proposed network. The learning rate is halved if the loss value remains unchanged for 20 epochs. The minimum learning rate is set to 0.0001. Additionally, we use a batch size of 1024 and an epoch size of 500 in subsequent experiments. For the downlink channel, we set the transmit power $P=46$ dBm and the noise power $\sigma^2_{dl}=-106$ dBm. We consider $K=2$ UEs, whose locations in the cell are randomly generated in each sample.


\subsection{Performance Comparison to Baseline Methods}
\label{A}
\begin{figure}[!t]
	\centering
	\includegraphics[width=3.3in]{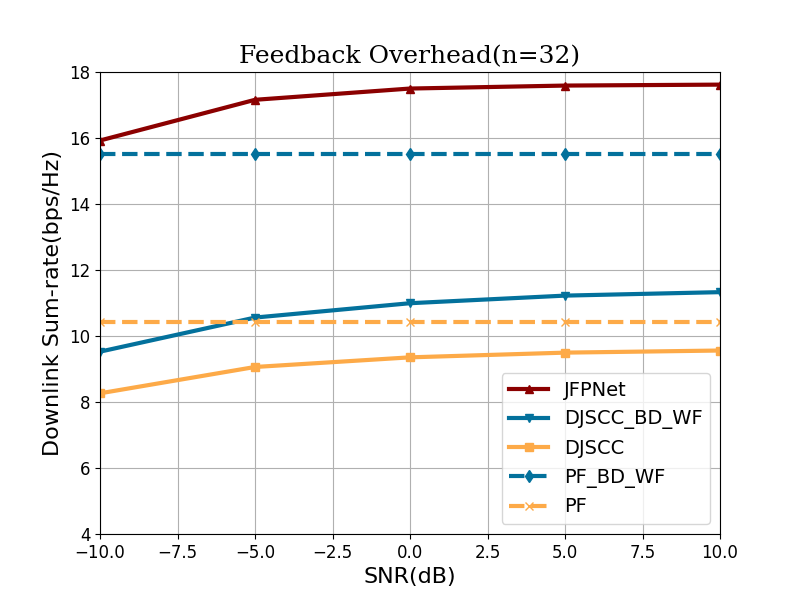}
	\caption{Performance comparison between baseline methods and AI-based methods.}
	\label{Ex1}
\end{figure}

In the experiment, the output size of the DJSCC encoder is fixed as $n=32$ and the downlink transmit power is also fixed. By adjusting the uplink transmit power, i.e., the uplink SNR, we evaluate the proposed method and the baseline methods. We train the proposed method under SNRs following a uniform distribution in the range of $[-10, 10]$dB. The scheme, that partial CSI eigenvectors recovered by the DJSCC are directly employed as the precoding matrix, is called ``DJSCC''. The ``DJSCC'' network was optimized with the reconstructed mean square error (MSE) during the training phase. Block diagonalization (BD) precoding based on partial CSI with water-filling (WF) power allocation is called ``DJSCC\_BD\_WF'' \cite{BD}. 
With the perfect feedback (PF) of the partial CSI, directly using feedback as the precoding matrix and using BD with WF algorithm are also considered as the reference, relevant to ``PF'' and ``PF\_BD\_WF''.

In Fig.~\ref{Ex1}, the JFPNet demonstrates a superior downlink sum-rate in comparison to the other methods. Compared with the ``DJSCC\_BD\_WF", JFPNet brings about a 6 bps/Hz performance gain in downlink sum-rate. Compared with ``DJSCC", the gain is expanded to 8 bps/Hz in the uplink $\mathrm{SNR \in \left[-5,10\right]}$ dB. As the uplink SNR increases, the performance of the conventional algorithm in the imperfect feedback case gradually approaches the upper bound in the PF case. At $\mathrm{SNR=10dB}$, a performance disparity of 0.87 bps/Hz exists between ``DJSCC" and ``PF". In contrast, there is a 4.2 bps/Hz performance differential between ``DJSCC\_BD\_WF" and ``PF\_BD\_WF", which shows that the performance of the BD algorithm and the WF algorithm is closely tied to the precision of the eigenvector matrix reconstruction by the DJSCC decoder. Even at $\mathrm{SNR=-10}$ dB, JFPNet still outperforms ``PF\_BD\_WF" by 0.4 bps/Hz, and this advantage becomes even more significant as the SNR increases.

\subsection{Ablation Study}
\label{B}
\begin{figure}[!t]
	\centering
	\includegraphics[width=3.3in]{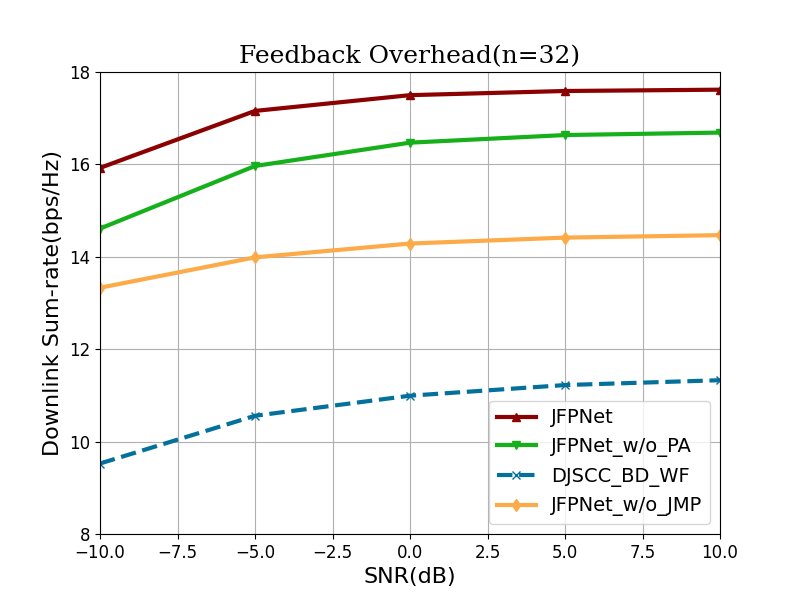}
	\caption{Ablation experiment. In JFPNet, three primary modules constitute its framework: the DJSCC module, the JMP module, and the PA module. }
	\label{Ex2}
\end{figure}

To better understand the proposed method and analyze the contribution ratio of the proposed modules, we compared the proposed method and its variation using an ablation approach. ``JFPNet w/o JMP" and ``JFPNet w/o PA" represent the JMP module and the PA module in the JFPNet are removed, respectively. The performance of ``DJSCC\_BD\_WF" is shown as the reference. As depicted in Figure~\ref{Ex2}, both the JMP module and the PA module play a significant role in improving the performance of the system. Compared with ``DJSCC\_BD\_WF", the JMP module and the PA module bring about 6 bps/Hz and 3 bps/Hz performance gain at $\rm SNR=0$ dB, respectively. This may be because the proposed method and its variation change the reconstruction objective of the DJSCC decoder from recovering compressed CSI to obtaining conducive matrices for precoding through the end-to-end training approach. This design facilitates implicit precoding reconstruction via the DJSCC decoder, leading to enhanced performance even in cases where precoding direction design is omitted. Moreover, the proposed JMP module contribute more than the proposed PA module in this case, e.g., about 2 bps/Hz at $\rm SNR=0$ dB. 

\subsection{Impact of Feedback Overhead}

\begin{figure}[!t]
	\centering
	\includegraphics[width=3.3in]{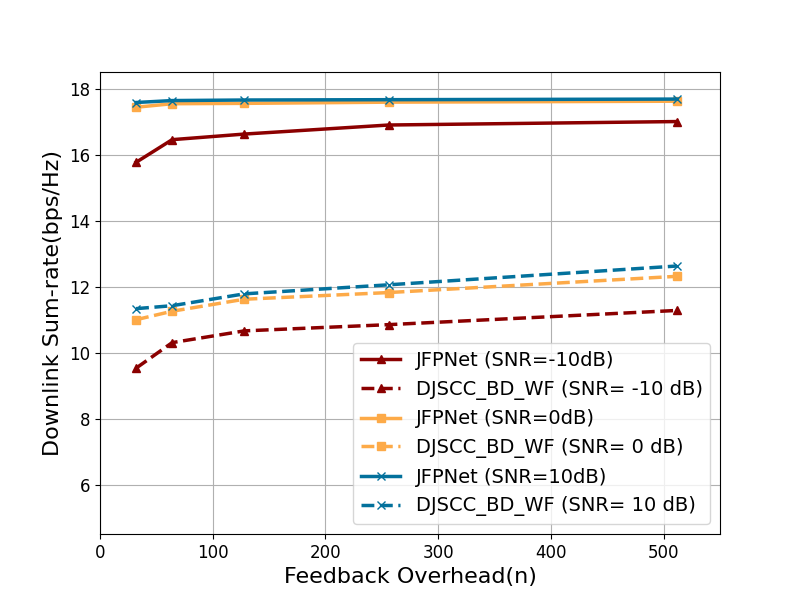}
	\caption{Network performance with different feedback overhead in different uplink SNR .}
	\label{Ex3}
\end{figure}
The performance under different feedback overhead $n$, i.e., 32, 64, 128, 256 and 512, is evaluated in this subsection. Consistent with experiments in Section \ref{A} and Section \ref{B}, the performance of the ``DJSCC\_BD\_WF'' is plotted as the reference. During the training stage, both the JFPNet and the DJSCC are trained with mixed uplink SNRs. The trained networks are evaluated at test SNRs of -10 dB, 0 dB and 10 dB. As shown in Fig.~\ref{Ex3}, the feedback overhead with a lower SNR, i.e., -10 dB, has greater impact on the downlink sum-rate than that with higher SNR, i.e., 10 dB. For instance, the downlink sum rate of the JFPNet with 512 feedback overhead is 1.24 bps/Hz higher than that with 32 feedback overhead at $\mathrm{SNR=-10}$ dB, while this gap at $\mathrm{SNR=10}$ dB is almost negligible. Moreover, the proposed JFPNet consistently outperforms traditional methods, particularly in the case of low feedback overhead. At $\mathrm{SNR=0}$ dB, with feedback overhead $n=32$ and $n=512$, the performance gain is 6.44 bps/Hz and 5.31 bps/Hz, respectively. 

\section{Conclusion}
\label{section4}

In this correspondence, we propose a joint CSI feedback and precoding method for the MU-MIMO-OFDM system. In the proposed JFPNet method, UEs feed eigenvector matrices with low feedback overhead via the DJSCC encoder, while the BS generate precoding vectors for multiuser by exploiting the DJSCC decoder, the JMP module and the PA module. To maximize the downlink sum-rate, the proposed network is trained in an end-to-end manner. Comparing with conventional methods, without increasing the feedback overhead, our method achieves a superior downlink sum-rate, especially under conditions of low feedback overhead and low SNR.


\bibliographystyle{IEEEtran}
\bibliography{myref}

\begin{thebibliography}{10}
\providecommand{\url}[1]{#1}
\csname url@samestyle\endcsname
\providecommand{\newblock}{\relax}
\providecommand{\bibinfo}[2]{#2}
\providecommand{\BIBentrySTDinterwordspacing}{\spaceskip=0pt\relax}
\providecommand{\BIBentryALTinterwordstretchfactor}{4}
\providecommand{\BIBentryALTinterwordspacing}{\spaceskip=\fontdimen2\font plus
\BIBentryALTinterwordstretchfactor\fontdimen3\font minus \fontdimen4\font\relax}
\providecommand{\BIBforeignlanguage}[2]{{%
\expandafter\ifx\csname l@#1\endcsname\relax
\typeout{** WARNING: IEEEtran.bst: No hyphenation pattern has been}%
\typeout{** loaded for the language `#1'. Using the pattern for}%
\typeout{** the default language instead.}%
\else
\language=\csname l@#1\endcsname
\fi
#2}}
\providecommand{\BIBdecl}{\relax}
\BIBdecl

\bibitem{CSI_overview}
J.~Guo, C.-K. Wen, S.~Jin, and G.~Y. Li, ``Overview of deep learning-based csi feedback in massive mimo systems,'' \emph{IEEE Transactions on Communications}, vol.~70, no.~12, pp. 8017--8045, 2022.

\bibitem{CsiNet}
C.-K. Wen, W.-T. Shih, and S.~Jin, ``Deep learning for massive mimo csi feedback,'' \emph{IEEE Wireless Communications Letters}, vol.~7, no.~5, pp. 748--751, 2018.

\bibitem{CsiNet_plus}
J.~Guo, C.-K. Wen, S.~Jin, and G.~Y. Li, ``Convolutional neural network-based multiple-rate compressive sensing for massive mimo csi feedback: Design, simulation, and analysis,'' \emph{IEEE Transactions on Wireless Communications}, vol.~19, no.~4, pp. 2827--2840, 2020.

\bibitem{CsiFormer}
X.~Bi, S.~Li, C.~Yu, and Y.~Zhang, ``A novel approach using convolutional transformer for massive mimo csi feedback,'' \emph{IEEE Wireless Communications Letters}, vol.~11, no.~5, pp. 1017--1021, 2022.

\bibitem{Implicit_CSI_Feedback}
M.~Chen, J.~Guo, C.-K. Wen, S.~Jin, G.~Y. Li, and A.~Yang, ``Deep learning-based implicit csi feedback in massive mimo,'' \emph{IEEE Transactions on Communications}, vol.~70, no.~2, pp. 935--950, 2021.

\bibitem{DJSCC}
J.~Xu, B.~Ai, N.~Wang, and W.~Chen, ``Deep joint source-channel coding for csi feedback: An end-to-end approach,'' \emph{IEEE Journal on Selected Areas in Communications}, vol.~41, no.~1, pp. 260--273, 2022.

\bibitem{MIMO_Pre}
M.~Zhang, J.~Gao, and C.~Zhong, ``A deep learning-based framework for low complexity multiuser mimo precoding design,'' \emph{IEEE Transactions on Wireless Communications}, vol.~21, no.~12, pp. 11\,193--11\,206, 2022.

\bibitem{MU_CSI_Pre}
K.~Wei, J.~Xu, W.~Xu, N.~Wang, and D.~Chen, ``Distributed neural precoding for hybrid mmwave mimo communications with limited feedback,'' \emph{IEEE Communications Letters}, vol.~26, no.~7, pp. 1568--1572, 2022.

\bibitem{MU_WeiYu}
F.~Sohrabi, K.~M. Attiah, and W.~Yu, ``Deep learning for distributed channel feedback and multiuser precoding in fdd massive mimo,'' \emph{IEEE Transactions on Wireless Communications}, vol.~20, no.~7, pp. 4044--4057, 2021.

\bibitem{MRFNet}
Z.~Hu, J.~Guo, G.~Liu, H.~Zheng, and J.~Xue, ``Mrfnet: A deep learning-based csi feedback approach of massive mimo systems,'' \emph{IEEE Communications Letters}, vol.~25, no.~10, pp. 3310--3314, 2021.

\bibitem{3GPP38214}
\vspace{0mm}3GPP, ``3rd generation partnership project; technical specification group radio access network; study on 3d channel model for lte (release 12),'' \emph{3GPP, Tech. Rep. 36.873 V12.7.0,}, 2020.

\bibitem{QuaDRiGa}
S.~Jaeckel, L.~Raschkowski, K.~Borner, and L.~Thiele, ``{QuaDRiGa}-quasi deterministic radio channel generator, user manual and documentation,'' \emph{Fraunhofer Heinrich Hertz Institute, Tech. Rep. v2.6.1,}, 2021.

\bibitem{3GPP38901}
3GPP, ``{5G} study on channel model for frequencies from 0.5 to 100 {GHz},'' {3rd Generation Partnership Project (3GPP)}, Technical Specification (TS) 38.901, 2020, version 16.1.0.

\bibitem{BD}
Q.~H. Spencer, A.~L. Swindlehurst, and M.~Haardt, ``Zero-forcing methods for downlink spatial multiplexing in multiuser mimo channels,'' \emph{IEEE transactions on signal processing}, vol.~52, no.~2, pp. 461--471, 2004.

\end{thebibliography}


 





\end{document}